\documentclass[conference,10pt]{IEEEtran}
\IEEEoverridecommandlockouts

\usepackage[T1]{fontenc}
\usepackage[inline]{enumitem}
\usepackage{cite}
\usepackage[pdftex]{graphicx}
\usepackage[table,xcdraw]{xcolor}
\graphicspath{figures}
\DeclareGraphicsExtensions{.pdf}
\usepackage{amsmath}
\usepackage{algorithmic}
\usepackage{array}
\usepackage[caption=false,font=footnotesize]{subfig}
\usepackage{bigstrut}
\usepackage{stfloats}
\usepackage{soul}
\usepackage{url}
\usepackage{fontawesome}
\usepackage{xspace}
\usepackage{tikz}
\usepackage{colortbl}
\usepackage{boldline}
\usepackage{hyperref}
\usepackage{caption}
\usepackage{enumitem}
\usetikzlibrary{shadows}
\usepackage[capitalise]{cleveref}
\usepackage[framemethod=tikz]{mdframed}
\usepackage[scale = 0.90]{plex-mono}
\usepackage{multirow}
\usepackage[newfloat,frozencache,cachedir=.]{minted}
\usepackage{tikz}
\usepackage{upquote}
\usepackage{nicefrac}
\usetikzlibrary{fit}
\definecolor{gray05}{gray}{0.95}
\definecolor{gray08}{gray}{0.92}
\definecolor{gray10}{gray}{0.90}
\definecolor{gray12}{gray}{0.88}
\definecolor{gray15}{gray}{0.85}
\definecolor{gray18}{gray}{0.82}
\definecolor{gray20}{gray}{0.80}
\definecolor{gray25}{gray}{0.75}
\definecolor{gray30}{gray}{0.70}
\definecolor{gray35}{gray}{0.65}
\definecolor{gray40}{gray}{0.60}
\definecolor{gray45}{gray}{0.55}
\definecolor{gray50}{gray}{0.50}
\definecolor{gray55}{gray}{0.45}
\definecolor{gray60}{gray}{0.40}
\definecolor{gray65}{gray}{0.35}
\definecolor{gray70}{gray}{0.30}
\definecolor{gray75}{gray}{0.25}
\definecolor{gray80}{gray}{0.20}
\definecolor{gray85}{gray}{0.15}
\definecolor{gray90}{gray}{0.10}
\definecolor{gray95}{gray}{0.05}
\definecolor{blue}{HTML}{0f62fe}
\definecolor{red}{RGB}{234,51,35}
\definecolor{ibm-blue-light}{HTML}{a6c8ff}
\hypersetup{
    colorlinks=true,
    linkcolor=blue,
    filecolor=magenta,      
    urlcolor=blue,
    citecolor=blue,
    pdftitle={Overleaf Example},
    pdfpagemode=FullScreen,
    }
\setminted{
    breaklines,
    frame=lines,
    bgcolor=white,
    framesep=1mm,
    baselinestretch=1.0,
    fontsize=\scriptsize, 
    breaklines, 
    breakanywhere, 
    linenos,
    numbersep=2pt,
    mathescape=true,
    style=xcode,
    escapeinside=||
}
\setitemize{noitemsep,topsep=0pt,parsep=0pt,partopsep=0pt,leftmargin=*, wide=0pt}
\setenumerate{noitemsep,topsep=0pt,parsep=0pt,partopsep=0pt,leftmargin=*, wide=0pt}
\newcommand{\subsect}[1]{{\color{blue}\small\P}\ref{sect:#1}\xspace}
\newcommand{\snippet}[1]{Listing~\ref{listing:#1}\xspace}
\newcommand\tool{\texttt{codellm-devkit}\xspace}
\newcommand\cldk{\textsc{cldk}\xspace}
\newcommand\bi{\begin{itemize}}
\newcommand\ei{\end{itemize}}
\newcommand\be{\begin{enumerate}}
\newcommand\ee{\end{enumerate}}
\newcommand{\inlinecode}[1]{\mintinline[fontsize=\footnotesize]{python}{#1}\xspace}
\newenvironment{code}{\captionsetup{type=listing}}{}
\SetupFloatingEnvironment{listing}{name=Listing}
\newcommand{\circled}[1]{%
  \begin{tikzpicture}[baseline=-0.6ex,
      circ/.style={shape=circle,draw,inner sep=1pt,fill=blue!90}]
    \node[circ] {\scriptsize\color{white}#1};
  \end{tikzpicture}%
}
\newcommand{\labeled}[2]{%
  \begin{tikzpicture}[baseline=-0.8ex,
      circ/.style={shape=circle,draw,inner sep=1pt,fill=blue!90}]
      \draw[dotted] (-#1ex, 0) -- (0, 0); 
    \node[circ] {\scriptsize\color{white}#2};
  \end{tikzpicture}%
}
\newcommand{\rtanglelabeled}[2]{%
  \begin{tikzpicture}[baseline=1ex,
      circ/.style={shape=circle,draw,inner sep=1pt,fill=blue!90}]
     \draw[dotted] (-#1ex, 1.5ex) -- (-#1ex, -0.3ex) -- (0, -0.3ex); % Draw an L-shaped dotted line leading to the circle if specified 
    \node[circ] {\scriptsize\color{white}#2};
  \end{tikzpicture}%
}
\newcommand{\inline}[1]{\mintinline[fontsize=\small]{python}{#1}}
\newmdenv[
    tikzsetting= {fill=blue!08},
    skipabove=0.5em,
    skipbelow=0.5em,
    linewidth=2pt,
    innerleftmargin=2pt,
    innerrightmargin=2pt,
    innertopmargin=2pt,
    innerbottommargin=3pt,
    linecolor=blue,
    roundcorner=2pt, 
    shadow=false,
    shadowsize=5pt,
    shadowcolor=blue!10
]{myshadowbox}
\newenvironment{lesson}
{\begin{myshadowbox} \textbf{Lesson:}\xspace}
{\end{myshadowbox}}

\newcommand{\hlpink}[1]{{\setlength\fboxsep{1pt}\colorbox{magenta!12}{#1}}}
\newcommand{\hlblue}[1]{{\setlength\fboxsep{1pt}\colorbox{blue!12}{#1}}}

\begin{document}
\title{Codellm-Devkit:~A Framework for Contextualizing Code LLMs with Program Analysis Insights}

% author names and affiliations
% use a multiple column layout for up to three different
% affiliations
% \author{\IEEEauthorblockN{Michael Shell}
% \IEEEauthorblockA{School of Electrical and\\Computer Engineering\\
% Georgia Institute of Technology\\
% Atlanta, Georgia 30332--0250\\
% Email: http://www.michaelshell.org/contact.html}
% \and
% \IEEEauthorblockN{Homer Simpson}
% \IEEEauthorblockA{Twentieth Century Fox\\
% Springfield, USA\\
% Email: homer@thesimpsons.com}
% \and
% \IEEEauthorblockN{James Kirk\\ and Montgomery Scott}
% \IEEEauthorblockA{Starfleet Academy\\
% San Francisco, California 96678-2391\\
% Telephone: (800) 555--1212\\
% Fax: (888) 555--1212}}

% conference papers do not typically use \thanks and this command
% is locked out in conference mode. If really needed, such as for
% the acknowledgment of grants, issue a \IEEEoverridecommandlockouts
% after \documentclass

% for over three affiliations, or if they all won't fit within the width
% of the page (and note that there is less available width in this regard for
% compsoc conferences compared to traditional conferences), use this
% alternative format:
% 
\author{\IEEEauthorblockN{Rahul Krishna\IEEEauthorrefmark{1}\IEEEauthorrefmark{2}\textsuperscript{1},
Rangeet Pan\IEEEauthorrefmark{1}\IEEEauthorrefmark{2}\textsuperscript{2},
Raju Pavuluri\textsuperscript{3}\IEEEauthorrefmark{2}, 
Srikanth Tamilselvam\textsuperscript{4}\IEEEauthorrefmark{3},
Maja Vukovic\textsuperscript{5}\IEEEauthorrefmark{2},
 and
Saurabh Sinha\textsuperscript{6}\IEEEauthorrefmark{2}}
\IEEEauthorblockA{\IEEEauthorrefmark{2}IBM Research, 
Yorktown Heights, NY 10598, USA.\\}
\IEEEauthorblockA{\IEEEauthorrefmark{3}IBM Research, 
Bangalore, KA 560045, India.\\}
\thanks{\scriptsize\IEEEauthorrefmark{1}~Equal Contribution}
\thanks{\scriptsize \textbf{Email}: \{\textsuperscript{$1$}rkrsn, \textsuperscript{$2$~}rangeet.pan\}@ibm.com, \{\textsuperscript{$3$}pavuluri, \textsuperscript{$5$}maja, \textsuperscript{$6$}sinhas\}@us.ibm.com, \textsuperscript{$4$}srikanth.tamilselvam@in.ibm.com}
}

% use for special paper notices
%\IEEEspecialpapernotice{(Invited Paper)}

% make the title area
\maketitle
% As a general rule, do not put math, special symbols or citations
% in the abstract
\begin{abstract}
Large Language Models for Code (or code LLMs) are increasingly gaining popularity and capabilities, offering a wide array of functionalities such as code completion, code generation, code summarization, test generation, code translation, and more. To leverage code LLMs to their full potential, developers must provide code-specific contextual information to the models. These are typically derived and distilled using program analysis tools. However, there exists a significant gap---these static analysis tools are often language-specific and come with a steep learning curve, making their effective use challenging. These tools are tailored to specific program languages, requiring developers to learn and manage multiple tools to cover various aspects of the their code base. Moreover, the complexity of configuring and integrating these tools into the existing development environments add an additional layer of difficulty. This challenge limits the potential benefits that could be gained from more widespread and effective use of static analysis in conjunction with LLMs. 

To address this challenge, we present \tool (hereafter, \cldk), an open-source library that significantly simplifies the process of performing program analysis at various levels of granularity for different programming languages to support code LLM use cases. As a Python library, \cldk offers developers an intuitive and user-friendly interface, making it incredibly easy to provide rich program analysis context to code LLMs. With this library, developers can effortlessly integrate detailed, code-specific insights that enhance the operational efficiency and effectiveness of LLMs in coding tasks. CLDK is available as an open-source library at \url{https://github.com/IBM/codellm-devkit}. 

\end{abstract}

\IEEEpeerreviewmaketitle
\section{Introduction}
\begin{figure}[htbp!]
    \centering
    \includegraphics[width=\linewidth]{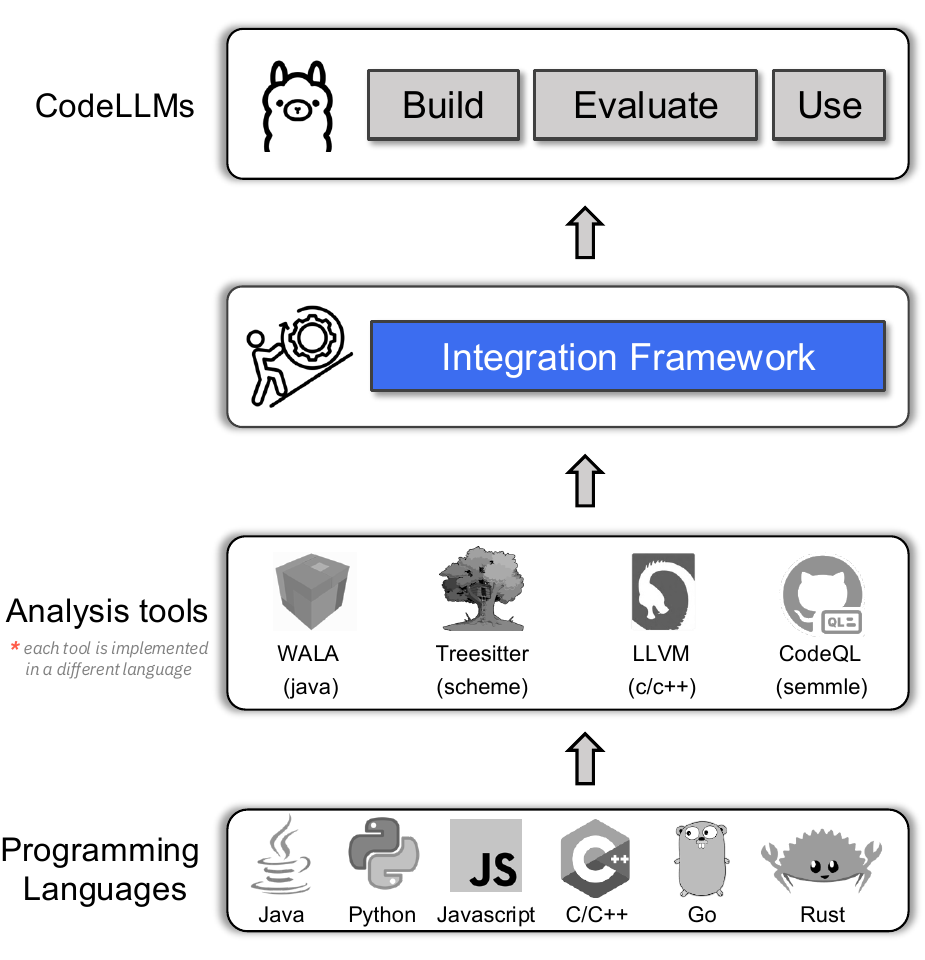}
    \caption{Landscape of LLMs based solutions for enterprise software systems.}
    \label{fig:introfig}
\end{figure}

The advent of Large Language Models (LLMs) has accelerated the use of insightful artificial intelligence in software development workflows. This has enabled a wide array of capabilities such as code completion\cite{completion1, completion2}, generation\cite{generation1, generation2, generation3}, summarization\cite{summarization1, summarization2, summarization3}, test generation\cite{test1, pan2024multi}, program repair\cite{repair2, repairrefactor1}, refactoring\cite{repairrefactor1}, and translation\cite{pan2024lost}. These language models have the potential to improve developer productivity and code quality significantly by automating routine tasks and providing intelligent insights.

To fully leverage the capabilities of LLMs for software engineering (SE) tasks, it is critical to provide them with rich, code-specific contextual information, which can be derived from program analysis tools that perform static and/or dynamic analysis on the codebases. These analysis tools can identify code patterns, dependencies, and potential issues that, when supplied to code LLMs, can enhance their contextual understanding of large codebases (beyond what was seen in their pre-training dataset) thereby impacting their overall performance, accuracy, and, as a consequence, their usefulness. 

Applying program analysis tools in the context of LLMs, however, presents several challenges: (1)~most existing static analysis tools are highly specialized for individual programming languages, each with their own expressive capabilities and idiosyncrasies, and further (2)~existing tools like WALA (for Java), LLVM (for Rust, and other C-family of languages), and CodeQL each adopt distinct internal representations, abstraction levels, and querying mechanisms making it hard to generalize across language families. As a result, enabling program analysis across multiple languages within a project introduces a number of unique challenges that arise not merely due to the diversity of languages, but also because of the differences in how these analysis tools model the code structure, the inherent semantic relationships, and their behavior. 

For developers, this means grappling with different APIs, data formats, and tooling conventions. The complexity arises from having to understand the specific design choices underpinning each tool—ranging from abstract syntax trees to intermediate representations and proprietary query languages—rather than simply managing tools for multiple languages. Further, integrating these tools into development environments often requires custom configurations and adaptations that further complicate the process, leading to significant overhead and limiting their usability.

\begin{figure*}[t!]
\begin{code}
\begin{minipage}[t]{0.33\linewidth}
\vspace{-0.5em}
\begin{minted}[linenos=false, baselinestretch=1, style=bw, framesep=3pt]{python}
query = f"""import java
    from Class c
    select c.getQualifiedName()
    """
with open(f"codeql.ql", "w") as f:
    |\hlblue{f.write(query)}|

result = |\hlblue{subprocess.run}|(f'codeql query run --database={APP_PATH} codeql.ql')

all_qualified_class_names = [cls.|\hlblue{split}|("|")[1].strip() 
  for cls in str(result.stdout.|\hlblue{strip()}|)
.split('\\n')[2:]]
\end{minted}
\vspace{-2.5em}
\vspace{0.5em}
\captionof{figure}{\small CodeQL requires numerous \hlblue{file i/o}, \hlblue{subprocess calls}, and \hlblue{string split/strip} to obtain an output.}
\label{fig:motivate-codeql}
\end{minipage}%
\end{code}
\begin{minipage}[t]{0.33\linewidth}
\begin{code}
\vspace{-0.5em}
\begin{minted}[linenos=false, baselinestretch=1, framesep=3pt]{python}


java_query=Language("java")
    .query(
        "(((class_declaration) @class))")

pyth_query=Language("python")
    .query(
        "(((class_definition) @class))")

go_query=Language("go")
    .query(
        "(((type_definition) @type_def))")

rust_query=Language("rust")
    .query("(((impl_item) @impl))")
\end{minted}
\vspace{-1.9em}
\captionof{figure}{\small Tree sitter has a different grammars\\ for similar abstractions in several languages}
\label{fig:motivate-treesitter}
\vspace{-3.1em}
\label{listing:cldk-core}
\end{code}
\end{minipage}~\begin{minipage}[t]{0.33\linewidth}
\begin{code}
\vspace{-.5em}
\begin{minted}[linenos=false, baselinestretch=1, framesep=3pt]{python}
from cldk import CLDK, Languages
from cldk.languages import Languages

cldk = CLDK(|\hlpink{language=Languages.Java})|
# The user only needs to change the 
# language, the APIs defined below will 
# remain the same.

analysis = cldk.analysis(Path(...))

# For languages (like go) which don't
# have classes, it's corresponding 
# equivalent (struct) will be returned.
classes = analysis.get_classes()
\end{minted}
\vspace{-1.9em}
\captionof{figure}{\small \cldk standarizes all API calls. The user only changes the \hlpink{language} parameter}
\vspace{-3.1em}
\label{fig:motivate-cldk}
\end{code}
\end{minipage}%
\end{figure*}

% \begin{figure*}[htp]
%     \centering
%     \includegraphics[width=\linewidth]{figures/motivation-color.pdf}
%     \caption{Canonical example of extracting all the classes in an application written in Java and Python using various static analysis tools like WALA~\cite{wala}, CodeQL~\cite{codeql}, Tree-sitter~\cite{treesitter}, and CLDK.}
%     \label{fig:motivation}
% \end{figure*}

This work offers \tool (hereafter, \cldk), an open-source Python-based library\footnote{Available at \href{https://github.com/IBM/codellm-devkit}{https://github.com/IBM/codellm-devkit}} designed to simplify the process of performing program analysis at various levels of granularity. \cldk provides an intuitive and user-friendly interface that abstracts the complexities of traditional static analysis tools. By supporting multiple programming languages and offering easy integration capabilities, \cldk enables developers to effortlessly extract rich program analysis context and supply it to code LLMs. By lowering the barrier to performing comprehensive program analysis, \cldk aims to empower developers to harness the full potential of code LLMs without the overhead of mastering multiple tools or dealing with complex integrations. We conjecture that \cldk will contribute significantly to the software engineering community by facilitating more efficient and effective development practices. Our contributions in this paper are as follows:

\begin{itemize}
   
    \item[$\circ$] \textbf{Development of \cldk.} We introduce \cldk, detailing its architecture, features, and how it addresses the identified challenges by providing a unified and simplified approach to program analysis.
   
    \item[$\circ$] \textbf{Integration with LLMs.} We demonstrate how \cldk can be seamlessly integrated with code LLMs to enhance various coding tasks, such as code generation and repair, by providing enriched contextual information.

   \item[$\circ$] \textbf{Developer Survey.} We interviewed 15 researchers and developers at IBM to understand their code LLM-based use cases and challenges that they face while using various program analysis and prompting tools. We also studied how \cldk can help ease their code LLM journey.
       
    \item[$\circ$] \textbf{Evaluation and case studies.} We evaluate the effectiveness of \cldk through real-world case studies and experiments that showcase its impact on improving the performance of code LLMs in real-world scenarios and improving research and developer productivity.
    
\end{itemize}

% \cldk is available as an open-source library at , and we invite the community to contribute, collaborate, and further advance this tool.

% The rest of this paper is structured as follows: in \sect{motivation}, we present the rationale behind \cldk, supported by a systematic literature survey and a preliminary user survey that highlights the needs and challenges faced by developers when working with language models for software engineering tasks. In \sect{design-principles}, we outline the architectural design of CLDK and we discuss the various levels of analysis, the modalities of interaction with large language models (LLMs), and techniques like retrieval and  augmentation (RAG) to enrich the interaction with LLMs. We also discuss how interoperability and extensibility considerations helped motivate the design pf CLDK, with particular attention to backend integration with systems like CodeQL, WALA, etc. The potential academic use cases of \cldk are discussed in \sect{academic-usecases}, and in \sect{industrial-usecases} we elaborate on the application to unit-test generation and code explanation offered in the IBM watsonx\textsuperscript{TM} Code Assistant suite of products. We then explore some of the lessons learned on \cldk’s potential impact on developer productivity in \sect{developer-productivity}. \sect{threats-to-validity} addresses possible limitations and we offer some concluding remarks in \sect{conclusion}.

\section{Motivation}
\label{sect:motivation}

The construction of \cldk was motivated by a pressing need to create a unified framework that integrates various tools required for building LLM-specific code use cases, while providing an abstraction layer that supports multiple programming languages. Performing analysis for different programming languages often requires understanding the intricate details of various program analysis tools, each of which can have a steep learning curve. Furthermore, these tools vary in their level of support for different languages, and extending analysis to additional languages often demands significant extra effort from developers.

To highlight these challenges, \cref{fig:motivate-codeql,fig:motivate-treesitter} demonstrate how different program analysis frameworks require vastly different approaches to accomplish the same task---that of having to retrieve all the class names from an application. Using CodeQL CLI~\cite{codeql}, the user needs to create a semmle query and save it to disk, then run it using the \texttt{subprocess} call, and then clean up the pipe-separated text output to extract the desired results. This process of performing multiple file i/o operations and execution of subprocess calls (which may require previlidge access) can be unscalable and may pose security threats. Further, performing string splits and strip can be challenging and error prone. Alternatively, with Tree-sitter~\cite{treesitter} (see~\cref{fig:motivate-treesitter}), simpler queries can be written, and the Python library can be used to parse the raw AST to retrieve class names. However, the developer needs to learn the grammar for each language, and construct new queries that leverage the language grammar. This has a steep learning curve and may not scale in practice.

\Cref{fig:motivate-cldk} illustrate how \cldk simplifies this process considerably---the developer just needs to instantiate a \cldk object (with the correct \hlpink{language} parameter), point it to the application, and call the \texttt{get\_classes} API to extract all class names from the application.

% Another challenge arises when extending the same analysis to a different programming language, such as Python. WALA, for instance, only supports Python up to version 2.7 due to its use of Jython~\cite{}, which poses problems for analyzing Python 3 code. In CodeQL, the user must modify the query imports, and with Tree-sitter, the query needs adjustment because node names differ between languages. In contrast, with \cldk, the user can simply switch the programming language to "Python," and the same analysis will work seamlessly, eliminating the need for additional knowledge about the backend implementation and greatly simplifying the process.
\section{Developer Survey on Tool Needs}
\label{sect:developer-productivity}

\begin{table}[]
\centering
\caption{Survey participants. Exp: Experience (in years).}
\label{tb:cohort}
\resizebox{\linewidth}{!}{\begin{tabular}{V{2}lV{2}lV{2}cV{2}cV{2}}
\hlineB{2}
\multicolumn{1}{V{2}cV{2}}{\textbf{\bfseries\color{blue}ID}} & \multicolumn{1}{cV{2}}{\textsf{\bfseries\color{blue}Designation}}                                                               & \textsf{\bfseries\color{blue}Division} & \textsf{\bfseries\color{blue}Exp.} \bigstrut\\ \hlineB{2}
\multicolumn{1}{c}{}&\multicolumn{1}{c}{}&\multicolumn{1}{c}{}&\multicolumn{1}{c}{}\bigstrut\\[-1.4em]\hlineB{2}

E1                             & Software Engineer                                                                                       & Engineer      & 5 \bigstrut\\%
%I3%
E2                              & \begin{tabular}[c]{@{}l@{}}Senior Technical Staff Member\end{tabular}         & Engineer      & 8             \bigstrut\\
E3                             & Product Manager~$^\text{\color{blue}M}$                                                                             & Engineer      & 10            \bigstrut\\
E4                              & \begin{tabular}[c]{@{}l@{}}Senior Technical Staff Member~$^\text{\color{blue}M}$\end{tabular} & Engineer      & 15            \bigstrut\\
E5                             & Senior Technical Staff Member~$^\text{\color{blue}M}$                                                               & Engineer      & 18            \bigstrut\\
E6                             & Senior Principal Software Engineer~$^\text{\color{blue}M}$                                                           & Engineer      & 25            \bigstrut\\
E7                              & Senior Technical Staff Member~$^\text{\color{blue}M}$                                                               & Engineer      & 26            \bigstrut\\
\hlineB{2}
\multicolumn{1}{c}{}&\multicolumn{1}{c}{}&\multicolumn{1}{c}{}&\multicolumn{1}{c}{}\bigstrut\\[-1.4em]\hlineB{2}

%I9%
R1                              & Research Engineer                                                                                       & Research      & 3             \bigstrut\\
R2                             & Research Engineer                                                                                       & Research      & 9             \bigstrut\\
R3                             & Senior Technical Staff Member~$^\text{\color{blue}M}$                                                               & Research      & 10            \bigstrut\\
R4                              & Senior Researcher                                                                                       & Research      &  15             \bigstrut\\
R5                              & Principal Research Scientist                                                                            & Research      & 20            \bigstrut\\
R6                              & Research Staff Member                                                                                   & Research      & 20            \bigstrut\\
R7                              & Senior Technical Staff Member                                                                           & Research      & 20            \bigstrut\\
R8                              & Principal Research Staff Member~$^\text{\color{blue}M}$                                                              & Research      & 30            \bigstrut\\\hlineB{2}
\multicolumn{4}{l}{$_{\color{blue}\text{M}{\text{\color{black}: Manager.}}}$}
\end{tabular}}
\label{tb:participants}
\end{table}

\begin{table*}[t!]
\caption{Developer survey questions.}
\label{tb:dev-survey}
\centering
\resizebox{\linewidth}{!}{
\begin{tabular}{V{2}cV{2}p{4in}V{2}lV{2}}
\hlineB{2}
\rowcolor[HTML]{FFFFFF} 
 \cellcolor[HTML]{FFFFFF}{\textbf{\textsf{\color{blue}Category}}}                                                                  & \multicolumn{1}{c}{\cellcolor[HTML]{FFFFFF}{\textbf{\textsf{\color{blue}Question}}}}                                                                                                                  & \multicolumn{1}{V{2}cV{2}}{\cellcolor[HTML]{FFFFFF}{\textbf{\textsf{\color{blue}Format}}}}                                                                                                                                   \bigstrut\\\hlineB{2}
 \multicolumn{1}{l}{}                                                                  & \multicolumn{1}{l}{}                                                                          & \multicolumn{1}{l}{}                                                                                                                                        \bigstrut\\[-1.4em]\hlineB{2}

\rowcolor[HTML]{FFFFFF} 
 \cellcolor[HTML]{FFFFFF}                                                                  & \textbf{A1.}~What is your job title?                                                                                                                 & \cellcolor[HTML]{FFFFFF}Open-ended                                                                                                                                   \bigstrut\\

 \multirow{-2.5}{*}{\cellcolor[HTML]{FFFFFF}\begin{tabular}{c}
Role and Experience
 \end{tabular}}                                                                   & \textbf{A2.}~How many years of software engineering experience do you have?                                                                          & \cellcolor[HTML]{FFFFFF}Numeric                                                                                                                                                \bigstrut\\\hlineB{2}
 \multicolumn{1}{l}{}                                                                  & \multicolumn{1}{l}{}                                                                          & \multicolumn{1}{l}{}                                                                                                                                        \bigstrut\\[-1.4em]\hlineB{2}
\rowcolor[HTML]{FFFFFF} 
 \cellcolor[HTML]{FFFFFF}                                                                  & \textbf{B1.}~How would you rate your experience with Code LLMs ?                                                               & MCQ                                                                                                                                              \bigstrut\\
\rowcolor[HTML]{FFFFFF} 
 \cellcolor[HTML]{FFFFFF}                                                                  & \textbf{B2.}~What are your primary use cases for Code LLMs?                                                                                          & MCQ                                                                                                                                        \bigstrut\\
\rowcolor[HTML]{FFFFFF} 
 \cellcolor[HTML]{FFFFFF}                                                                  & \textbf{B3.}~What programming language(s) do you use to build CodeLLM applications?                                                         & Open-ended                                                                                                                                                                     \bigstrut\\
 
\rowcolor[HTML]{FFFFFF} \cellcolor[HTML]{FFFFFF}
              & \textbf{B4.}~What is the target programming language(s) you use CodeLLMs for?                                                                      & Open-ended                                                                                                                                                                     \bigstrut\\
\rowcolor[HTML]{FFFFFF} 
 \cellcolor[HTML]{FFFFFF}                                                                  & \textbf{B5.}~Does your CodeLLM usecase require additional program analysis?                                                                          & MCQ                                                                                                                                        \bigstrut\\
\rowcolor[HTML]{FFFFFF} 
 \cellcolor[HTML]{FFFFFF}                                                                  & \textbf{B6.}~Do you use any program analysis tools?                                                                                                  & MCQ                                                                                                                                                 \bigstrut\\
\rowcolor[HTML]{FFFFFF} 
 \cellcolor[HTML]{FFFFFF}                                                                  & \textbf{B7.}~On average, how much time do you spend on prototyping and building your Code LLM use cases?                                             & MCQ                                                                                                                                                 \bigstrut\\
\rowcolor[HTML]{FFFFFF} 
 \multirow{-14}{*}{\cellcolor[HTML]{FFFFFF}\begin{tabular}{c}
Background of\\LLM use cases
 \end{tabular}}                                                                  & \textbf{B8.}~What fraction of the time do you typically spend obtaining program analysis insights for your projects?                                 & MCQ                                                                                                                                                 \bigstrut\\\hlineB{2}
\rowcolor[HTML]{FFFFFF} 
 \multicolumn{1}{l}{}                                                                  & \multicolumn{1}{l}{}                                                                          & \multicolumn{1}{l}{}                                                                                                                                        \bigstrut\\[-1.4em]\hlineB{2}

\rowcolor[HTML]{FFFFFF} 
 \cellcolor[HTML]{FFFFFF}                                    & \textbf{C1.}~What are the pros/cons of the tool(s) used? Please explain.                                                                             & Open-ended                                                                                                                                                                     \bigstrut\\

\rowcolor[HTML]{FFFFFF} 
 \multicolumn{1}{V{2}lV{2}}{\cellcolor[HTML]{FFFFFF}}                                              & \textbf{C2.}~What features or capabilities do you find missing in current program analysis tools?                                                    & Open-ended                                                                                                                                                                     \bigstrut\\
\rowcolor[HTML]{FFFFFF} 
 \multirow{2}{*}{\begin{tabular}{c}Need for Better/Easier\\Program Analysis Tools\end{tabular}}                                              & \textbf{C3.}~How effective are your current tools when working with new programming languages or applications?                                       & Likert Scale                                                                                                                                                              \bigstrut\\
\rowcolor[HTML]{FFFFFF} \multicolumn{1}{V{2}lV{2}}{\cellcolor[HTML]{FFFFFF}}                                              & \textbf{C4.}~How would you rate the learning curve of your program analysis tools when applied to new languages or applications?                     & Likert Scale                                                                                                                                                              \bigstrut\\

\rowcolor[HTML]{FFFFFF} 
  & \textbf{C5.}~Do you feel there is a need for better or easier-to-use program analysis tool for LLM applications? Please explain.                     & MCQ      
\bigstrut\\\hlineB{2}
\rowcolor[HTML]{FFFFFF} 
 \multicolumn{1}{l}{}                                                                  & \multicolumn{1}{l}{}                                                                          & \multicolumn{1}{l}{}                                                                                                                                        \bigstrut\\[-1.4em]\hlineB{2}

\cellcolor[HTML]{FFFFFF}                                                                  & \cellcolor[HTML]{FFFFFF}\textbf{D1.}~Do you use any prompting tool? If yes, which tools and what are the ideal features you would look for.          & \cellcolor[HTML]{FFFFFF}Open-ended                                                  \bigstrut\\
\cellcolor[HTML]{FFFFFF}                                                        \multirow{-2}{*}{\cellcolor[HTML]{FFFFFF}Prompting and RAG}          & \cellcolor[HTML]{FFFFFF}\textbf{D2.}~Do you use any tool for RAG? If yes, use any tool and what are the ideal features you would look for.           & \cellcolor[HTML]{FFFFFF}Open-ended                                                                                    \bigstrut\\\hlineB{2}
\rowcolor[HTML]{FFFFFF} 
\end{tabular}}
\end{table*}

In this section, we describe a study in which we learn the overall experience of IBM developers and researchers in the building, using, and interacting with code LLMs and their need for systematic program analysis for their use cases. We conducted an interview of 15 IBM Research employees, who have been an integral part of the organization's AI for Code initiative. Our main objective is to understand (a) the overall experience of code LLM, (b) various use cases of code LLMs, (c) experience with existing program analysis tools, (d) experience with prompting and other aspect of code LLM and various available tools, and (e) finally we investigate a developers' and researchers' wishlist on a tool that they would prefer to use in their workflow. 
% To that end, first, we discuss the interview design including questionnaires and then discuss participants' responses.

\noindent\textbf{Cohort Selection.~}Our survey involved 15 participants who actively build and/or use CodeLLMs in enterprise pipelines at IBM. The survey cohort represents a balance between researchers (8 participants) and engineers (7 participants) to gather diverse insights with varying degrees of experience (with a median of 15 years of SE and 17\textonehalf\xspace years of Research experience see~\cref{tb:cohort}). We interviewed each participant over a 30-minute call and recorded the session. Participants for whom we exhausted our allocated meeting time or for whom additional information was needed before answering our questions, we followed-up with additional sessions. 

\noindent\textbf{Questionnaire Design.~}We divided the questionnaire into four categories -- (a) The roles and experiences of the participants, (b) background and demographics of the participants, their exposure to Code LLMs, and their Code LLM use cases, (c) challenges pertaining to the use of program analysis and construction code LLM-based solutions; (d) prompting and LLM features, where we query the participants' use of prompting, RAG, agents and other framework.%, and lastly (d) The participants to expand on their ideal tool to be used with (and for) CodeLLMs.

\subsection{Interview results}

\subsubsection{Background of solutions with code LLM}
\begin{figure}[htp]
    \centering
    \includegraphics[width=\linewidth]{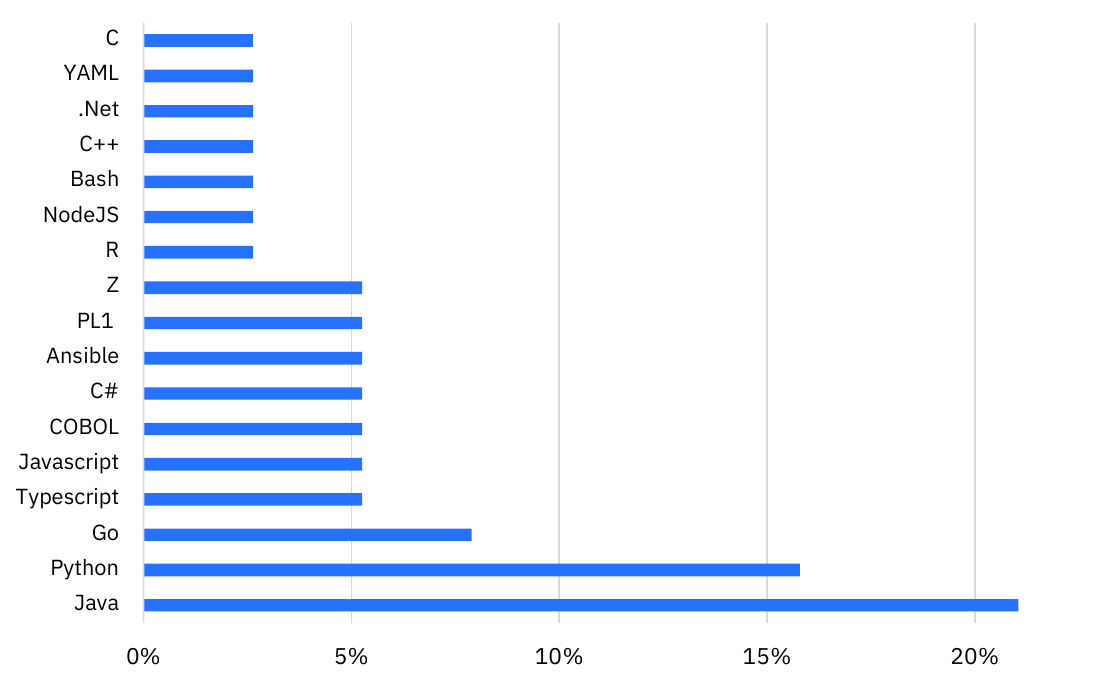}
    \caption{Target programming languages for LLM-assisted tasks at IBM.}
    \label{fig:interviewpl}
\end{figure}

Out of 15 participants, we observed a wide variety of designation (\textbf{A1}) and a large distribution of SE experience (A2) with a minimum of 3 years to a maximum of 30 years (see~\cref{tb:cohort}). In terms of their background in LLMs, we made the following key observations:
\bi
\item \textbf{Rating LLM experience (B1).}~A majority of the participants had either high (53\%) or medium (33\%) exposure to using LLM with a minority (14\%) had low exposure to LLM-based use cases, respectively. 
\item \textbf{Primary LLM use-case (B2).~} The majority or our participants mentioned use cases such as code generation, code explanation, code repair, test generation, code refactoring, and application modernization. Two participants indicated their primary focus was \textit{developing new prompting languages}. 
\item \textbf{The lingua-franca of LLM development  (B3).~}Nearly all participants noted that they use Python to develop LLM-based solutions, while one participant mentioned using TypeScript, primarily due to having to build plugins for VS Code.
\item \textbf{The target language for LLM development (B4).~} Most code LLM use cases, according to our participants, were focused on Java followed closely by Python, Go, and Javascript/Typescript. Some participants also used LLMs for legacy languages such as COBOL and PL1. The overall distribution is shown in~\cref{fig:interviewpl}.

\item \textbf{Use of program analysis (B5).~}All but two participants (R1 and E2) agreed that that they use program analysis in their code LLM pipeline. Of the two exceptions, R1 answered that their use case is surrounding taking LLM output off-the-shelf whereas E1 works on building an evaluation pipeline for code LLM (which does not warrant additional program analysis). 

\item \textbf{Program analysis tool use (B6).~}Of the participants that answered affirmatively to the above, more than 60\% of them have used various open-source tools like Tree-sitter, WALA, JavaParser~\cite{javaparser}, Soot~\cite{soot}, LSP, etc,. The remaining participants used/constructed in-house tools. 

\item \textbf{Time to prototyping vs time to productionizing (B7).~} Majority of our interview participants agreed that a prototype typically takes a week to a month, whereas a full production take anywhere from 6 months to a year. 
\item \textbf{Time spent obtaining program analysis (B8).~} Most of our participants suggested that they spend a medium amount of their time obtaining analysis insights. Some engineers suggested delegating the analysis to subject matter experts on program analysis to alleviate the learning curve.

\ei
\begin{lesson}
Most participants had medium to high LLM experience prototyping and building CodeLLM based applications. Python was their dominant language for development, with Java as their main target language. \textbf{\textit{Most used program analysis tools, either open-source or custom-built, and spent a significant amount of time obtaining program analysis insights}} highlighting that program analysis is both critical and time-intensive.
\end{lesson}
\subsubsection{Need for Better/Easier Program Analysis Tools}
In these questions, we attempt to understand the key challenges users face when they program analysis in their solution. Overall, we identified a few areas that could be improved for the existing program analysis approaches.
\begin{itemize}
    \item \textbf{Program Analysis Depth (C1 and C2).} A key concern raised was the availability of various tools that can perform program analysis at different levels of complexity. Participants unanimously agreed that many LLM-based use cases require lighter analysis, which can be efficiently handled by AST-based tools like Tree-sitter and Antlr. However, more complex analyses necessitate the use of advanced tools such as WALA, Soot, or LLVM.
    \item \textbf{Support for multiple programming languages (C3).} Our participants rated this aspect at an average of 2 (not very effective) on a scale of 1 (not at all effective) to 5 (very effective). A major concern was the scattered support for various languages often means that extending support for another language requires switching tools and rewriting analysis code from scratch, which they suggested hampered their productivity. Participants agreed unanimously that a tool with comprehensive support for multiple languages would significantly enhance productivity and streamline their efforts.
    \item \textbf{Learning curve (C4).} On average, all participants rated the learning curve as high (4) or very high (5) with the lowest rating being 1 (very low). They indicated that each tool presents its own learning challenges. For instance, participants using WALA acknowledged that, despite its powerful capabilities for in-depth program analysis, it has a steep learning curve. 
    % Figure~\ref{fig:motivation} illustrates this point, showing how fetching all classes in a Java application requires substantial coding effort and a solid understanding of program analysis. Additionally, the need to learn a new tool when extending support to another language further complicates the process.
    
    \item \textbf{Need for better tools (C5).} When posed with this question, participants bemoaned the lack of inconsistency among different tools, particularly in terms of analysis precision, flexibility for custom analysis, output format, required skill level, and verification of analysis accuracy. These challenges overlap with previous discussions on analysis depth and the skills needed to integrate these tools into a workflow. 
    With regards to flexibility, most participants found tree-sitter favourable due to its straightforward query-based interfaces. One participant noted that, \textit{``Tree-sitter, being a collection of grammars rather than a typical analysis tool, offers great flexibility''}. However, many participant lamented the need for extensive extensive post-processing to consume the generated output from tree sitter. All participants indicated that a good documentation is paramount in choosing the tool (with Tree-sitter and CodeQL being favored for this aspect).
\end{itemize}
\begin{lesson}
    Participants highlighted a need for better/easy-to-use program analysis tools. They lamented the lack of comprehensive multi-language support and the inconsistent capabilities across tools, forcing them to switch between tools and rewrite code. There was a clear demand for more flexible, well-documented tools that can handle multiple languages and provide seamless analysis with minimal effort.
\end{lesson}
\subsubsection{Support for prompting and other LLM features} In these questions, we attempt to understand the use of various tools for prompting, RAG, and agentic workflow.
\begin{itemize}
    \item \textbf{Use of prompting tools (D1).}~While some of participants indicated that they use prompting tools such as LangChain, BedRock, and PDL for tasks like code generation. Most indicated they use in-house tools, which work on a trial-and-error basis or in a free-form manner. Participants expressed a need for prompting tools with improved flexibility, easier integration, and support for various models. A few preferred tools that aid in building prompts empirically based on experience.
    
    \item \textbf{Use of RAG tools (D2).}~Participants were split between using in-house solutions and established tools like ElasticSearch and MPNet for retrieval-augmented generation (RAG). Some participants highlighted the need for faster search and support for diverse embedding models. While others indicated that they value tools that simplify encoding, search, and retrieval processes. Flexibility in supporting various models and embeddings, as well as better performance for search and retrieval, were seen as key features.
\end{itemize}
\begin{lesson}
    Participants highlighted the need for more flexible and efficient prompting and RAG tools to improve their workflow. Many rely on in-house or hit-and-trial methods, with a strong desire for tools that support multiple models, provide faster search, and simplify encoding and retrieval processes. Better integration and reduced learning curves for these tools would significantly enhance productivity.
\end{lesson}
% \subsubsection{Ideal code LLM tool} 

\section{Academic Survey on Tool Needs}
\label{sect:academic-usecases}
In view of increased LLMs adoption in SE, we explore the extent to which academia incorporates program analysis and for which specific use cases.  
For this, we conducted a systematic review of relevant academic literature by mining papers from top-tier venues in SE and AI. We categorized the papers by their software engineering tasks and manually analyzed each study to understand the role and application of program analysis within these contexts. This section outlines our key insights from this investigation.
\subsection{Dataset collection} To collect the list of papers, first, we filtered all the works since the beginning of 2023, when the vast usage of LLM has been started. We identified papers published in top-tier venues like ICSE, FSE, ASE, ISSTA, AAAI, NeurIPS, ICML, etc,. Also, we filtered paper by searching for keywords specific to various widely used program analysis frameworks, such as ``treesitter'', ``tree-sitter'', ``codeql'', ``wala'', etc,. Upon applying all these filtering criteria, we found 34 papers. Then, we manually went through to (a) categorize the work into various SE tasks, (b) target programming languages, (c) tool used for program analysis, and (d) verify the availability of the artifact. For categorizing the papers into SE tasks, we relied on an existing work~\cite{hou2023large}.

\subsection{Dataset insights}
\begin{figure}[htp]
    \centering
    \includegraphics[width=\linewidth]{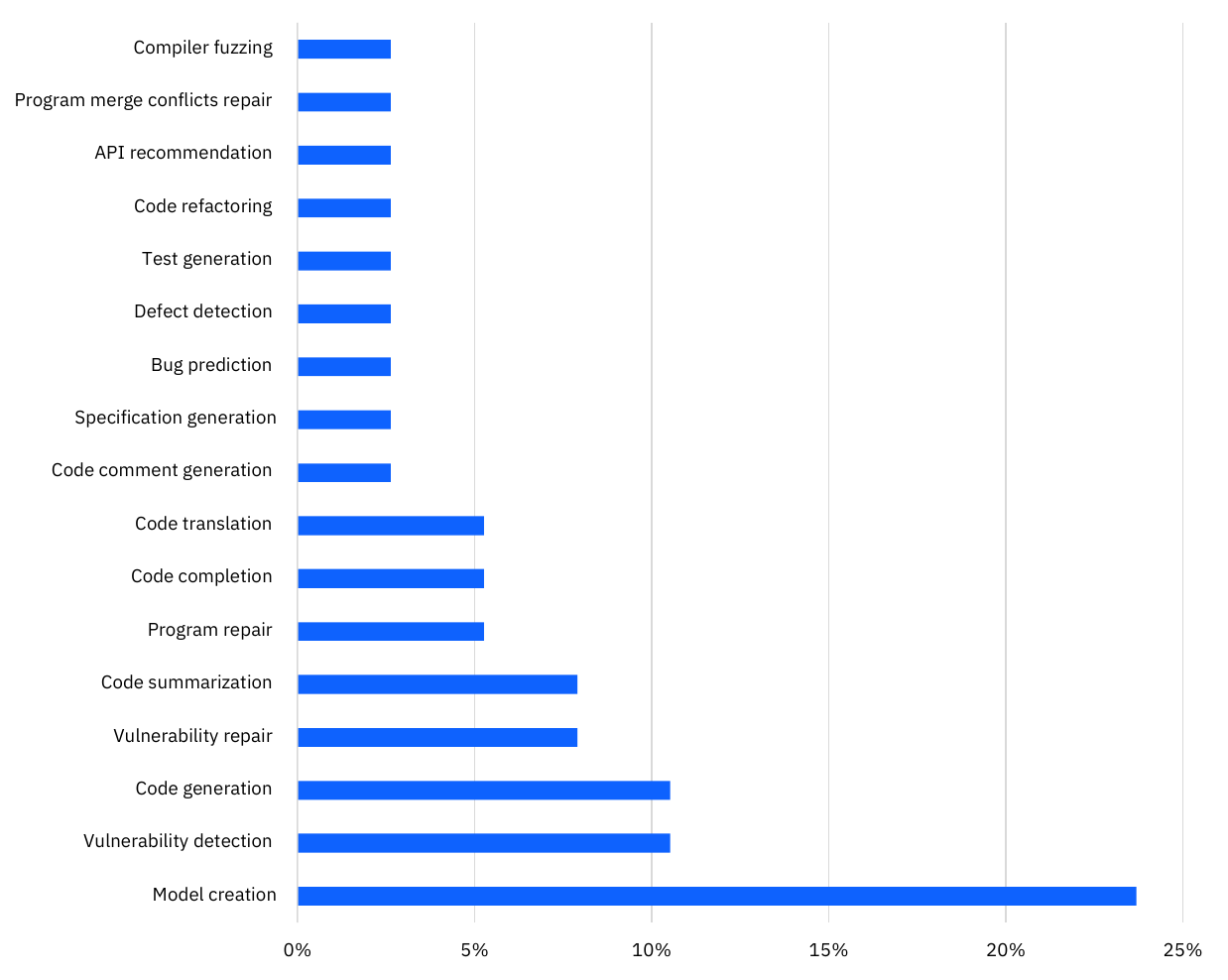}
    \caption{LLM-assisted SE tasks where program analysis has been an integral part.}
    \label{fig:setasks}
\end{figure}

\begin{figure}[htp]
    \centering
    \includegraphics[width=\linewidth]{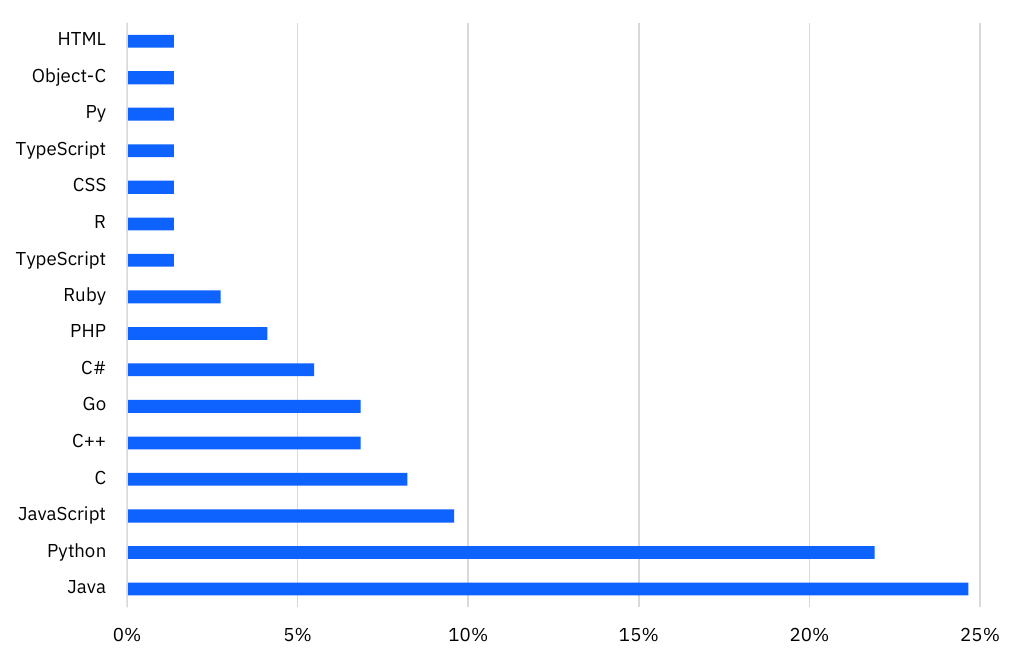}
    \caption{Target programming languages for LLM-assisted tasks based on manual investigation of academic works.}
    \label{fig:pl}
\end{figure}

We categorized the works related to code and LLM usage (depicted in Figure~\ref{fig:setasks} and found that most works are related to creating new models, vulnerability detection, code generation, code completion, etc. Also we looked into the programming languages the works are focused on. We found that the majority of the work is targeted for Java, Python, and JavaScript as described in Figure~\ref{fig:pl}. Also, a majority of the work is for C and C++, which are mostly related to vulnerability detection. Languages like JavaScript, Go, C\#, PHP are also being focus of several works. This has been a big motivation for the list of languages to be included in the list of supported programming languages for \cldk.
\begin{lesson}
    LLMs are used for a wide range of SE tasks supporting a variety of programming languages. Also, a significant portion of these academic works incorporate program analysis in their pipeline, making a potential candidate for \cldk.
\end{lesson}

\section{Design Principles}
\label{sect:design-principles}
\begin{figure}[tbp!]
    \centering
    \includegraphics[width=\linewidth]{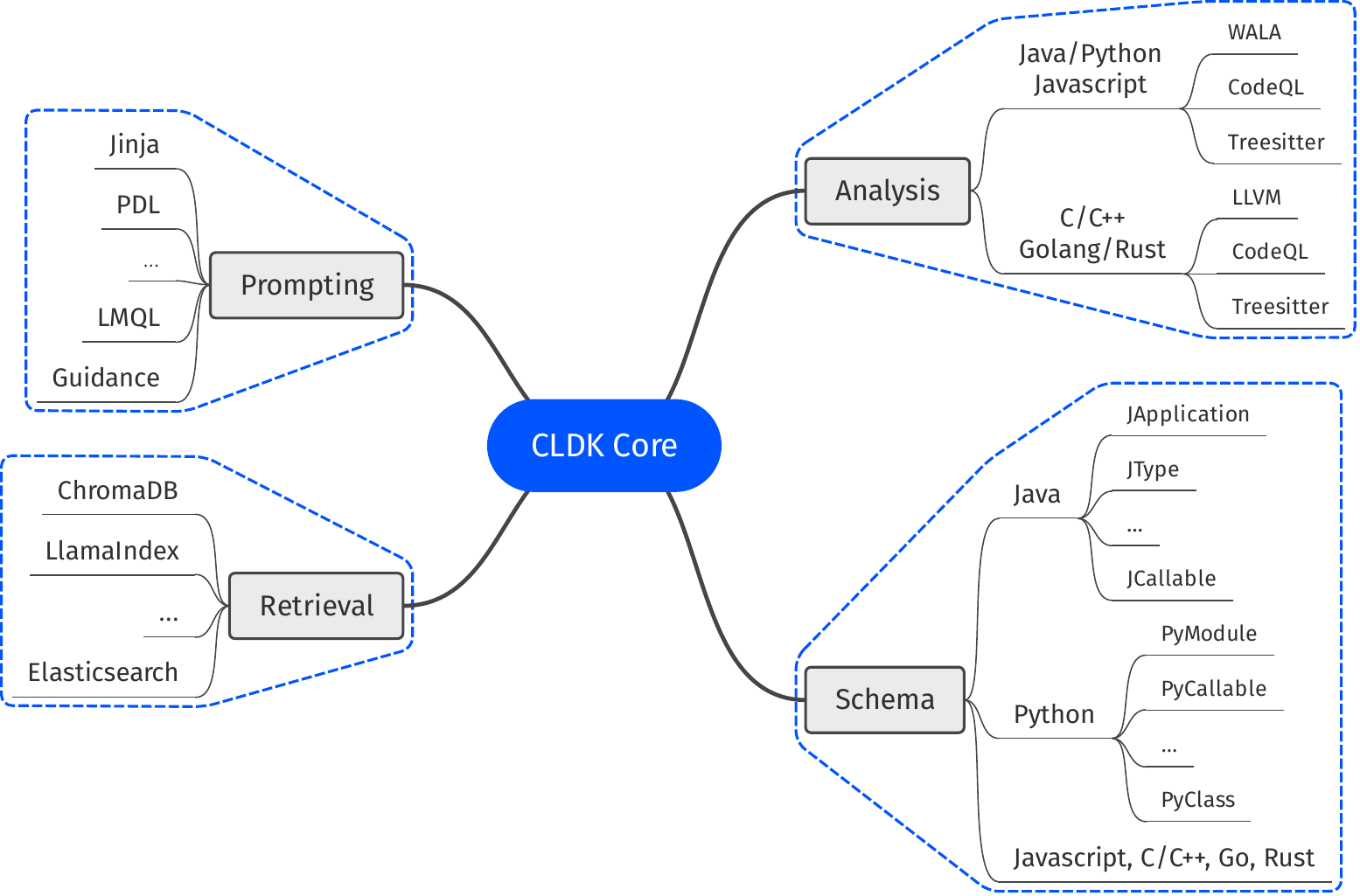}
    \caption{System overview of \protect\tool.}
    \label{fig:cldk-sys-overview}
\end{figure}

% Note to self--add the following details in this section:

% Start with a general statement about the overall architecture.
% Mention whether it's modular, monolithic, distributed, etc., and the main goal behind this structure (e.g., maintainability, scalability).

% Describe Key Components
% Briefly introduce the main components or modules.
% Explain the functional responsibilities of each component in one sentence.
% Explain Component Interactions

% Describe how the components interact, focusing on data flow or communication protocols.
% Mention any interfaces or APIs that enable this interaction.
% Provide Deployment Context

% Describe where the tool runs: deployment model (e.g., cloud, on-premises, client-server).
% Explain how components are distributed, if relevant.

% Highlight Design Patterns
% Mention any design patterns used in structuring the tool (e.g., MVC, event-driven).
% State why these patterns were chosen.

% Technology Stack and Rationale
% Briefly list the key technologies (languages, frameworks, tools).
% Explain the rationale for choosing these technologies.

% Scalability and Extensibility
% Address scaling mechanisms (horizontal or vertical).
% Mention points of extensibility for future growth.

% Challenges Addressed
% Conclude with a mention of significant challenges the architecture solves.
% State how the chosen architecture addresses these challenges.

In this section, we discuss the design principles underpinning \cldk (\tool). \cldk adopts a 
modular structure, where each module has a clearly defined functional responsibility such as analysis of programs, a schematic definition for each language, mechanisms to incorporate prompting and retrieval. 

The primary design objective of \cldk is to ensure:~1)~Clean separation of concerns; 2)~The ability to independently update and/or add extensions to each module; and 3)~Enhanced ability to test each module independently. \cref{fig:cldk-sys-overview} provides an overview of how \cldk operates and will be explained in the remainder of this section.

% In this section, we discuss the design principles underpinning \tool. \cldk adopts a fundamentally different approach
% from prior graph capture systems in PyTorch. Rather than
% trying to remove or replace Python, \cldk tries to
% work with CPython by just-in-time (JIT) compiling Python
% bytecode. \cldk is a Python bytecode to Python
% bytecode translator, where it extracts PyTorch operations
% from the original bytecode and replaces them with calls to
% compiled artifacts that fuse many PyTorch operations together. Figure 1 provides an overview of how \cldk4 operates and will be explained in the remainder of this section.

\subsection{Architectural Overview}
\label{sect:architectural}

In this section, we provide a detailed description of the core components of \cldk outlining their expected behavior and illustrating how they function together.

\subsubsection{CLDK core}
This module serves as an entrypoint to \tool. It provides the interface for interacting with the underlying functionalities. This module offers all the encapsulation necessary to configure \cldk, customize various options based on the programming languages, and to run analyses. A typical usage of \cldk involves instantiation with a target programming language. Next, we create an \texttt{analysis} object configured with language specific options to then invoke specific functions for analysis or processing as needed. 
An example usage for a Java project is shown in~\snippet{cldkcore}
\begin{figure}[!htbp]
\begin{code}
\vspace{-1em}
\begin{minted}{python}
cldk = |{\color{blue}\textbf{CLDK}}|(language="java")
j_analysis: |\hlblue{JavaAnalysis = cldk.analysis(}| |\labeled{24.5}{a}|
        project_path="path/to/project", 
        |\hlblue{\texttt{analysis\_engine=Engines.CodeQL}}, \labeled{26.8}{b}|
        **kwargs)  # Other optional analysis options.
\end{minted}
\vspace{-2.5em}
\captionof{listing}{A sample instantiation of \cldk for Java.}
\vspace{0.5em}
\label{listing:cldkcore}
\end{code}
\end{figure}

\inline{Analysis} module is one of the core components of \cldk that serves the functionality of examining the codebases, extracting useful information, and persisting them in the analysis scope by creating a robust and reusable schema based on the pydantic notation~\cite{pydantic} (discussed further in \subsect{data-model}). This module comprises of a language specific handler that offers several useful APIs that facilitate  user interaction with program analysis tools. For instance, in~\snippet{cldkcore}, the developer would first instantiate \cldk (\circled{a}) as the user invokes \inline{cldk.analysis(...)} for a java project, they obtain an object of type \hlblue{\small\texttt{JavaAnalysis}} which encapsulates all APIs which can offer analysis artifacts over the java project using bespoke getter methods. 
\begin{figure}[!t]
\begin{code}
\vspace{-0.5em}
\begin{minted}{python}
# Get the application call graph as a NetworkX 
call_graph: |\hlblue{nx.DiGraph}| = |j\_analysis.\hlblue{get\_call\_graph()}||\labeled{12}{a}|
                |\rtanglelabeled{55}{b}|
\end{minted}
\vspace{-3em}
\begin{minted}{python}
# Get a dictionary of all the methods in a class.
methods: |\hlblue{Dict[str, JCallable]} \labeled{38}{c}\vspace{0.25em}|
    = j_analysis.|\hlblue{get\_methods\_in\_class(}\labeled{28.7}{{\scriptsize{d}}}|
          qualified_class_name="foo.bar.ClassName")
\end{minted}
\vspace{-2.5em}
\captionof{listing}{Sample API calls over the Java application.}
\vspace{-0.5em}
\label{listing:cldk-api-use-1}
\end{code}
\end{figure}
From~\snippet{cldk-api-use-1}, we observe that: i) the {\small\texttt{JavaAnalysis}} object makes available convenient getter methods to access analysis artifacts such as call-graph (\circled{a}, in~\snippet{cldk-api-use-1}) and a enumeration of all the methods in a given class (\circled{d}, in~\snippet{cldk-api-use-1}), among several others; and ii) The returned artifacts are native python objects such as a {\small\texttt{networkx.DiGraph}} (\circled{b}, in~\snippet{cldk-api-use-1}) or a python dictionary of string and {\small\texttt{JCallable}} (\circled{c}, in~\snippet{cldk-api-use-1})---a native ``pydantic'' data class discussed in detail below.

\subsubsection{Program analysis Engines}

Program analysis engines provide the underlying analysis support for \cldk. They are invoked indirectly via the {\color{blue}\small\texttt{analysis\_engine}} argument. \cldk supports various program analysis backends, each designed to cater the needs of the various supported languages. For instance, LLVM~\cite{llvm} is used for C-like languages such as C, C++, Go, and Rust. Similarly, WALA~\cite{wala} is used for analyzing Java, Javascript, and (to a growing extent) Python. Both LLVM and WALA provide rigorous static analysis insights for their respective languages. In addition, \cldk offers support for CodeQL~\cite{codeql} to handle several language-agnostic analyses across a broad spectrum languages and framework. 

Program analysis requirements vary with the programming language. \cldk accommodates this variance by integrating multiple engines and analysis levels to provide a thorough yet consistent developer experience. This is achieved by performing a preliminary analysis with Tree-sitter~\cite{treesitter} (which serves as a parser capable of generating syntax trees for a variety of languages) and using this as a basis for more advanced levels of analysis (such as control flow and/or data flow analysis) using language specific tools like LLVM, CodeQL, or WALA. The modular nature of the analysis backends allows developers to customize the depth of the analysis depending on the needs of the project, ranging from simple syntactic checks to complex dependency analysis, thereby providing flexibility and extensibility for diverse use cases.
Further, \cldk is designed in a manner such that the developer is free to choose any backend they may see fit, and \cldk will guarantee a consistent output regardless of their choice.

% (as illustrated in~\cref{fig:introfig}, \circled{gray70}{2})

\subsubsection{Schema}
\label{sect:data-model}
% In this project, schema is an interconnected set of pydantic data model that encapsulates the various components of an application (written in one or more PLs). The abstraction will always contain a Application Base Class, which will contain fields for "Symbol Table", "Call Graph", "Local Control/Data Flow Graphs", and "System Dependency Graphs". Each of the above have further granular sub-divisions to accomodate "Files/Types/Classes/Modules/Structs/Traits" which inturn contain Dataclasses for "Callables", "Imports", "Fileds", etc. 

% Why do we have such a schema? What benefits does it confer when we are working on multiple languages? Why did we choose pydantic?

The schema provides a multi-lingual abstraction to standardize the representation of various components of an application, such as dependencies (imports), files, modules, classes, types, structs, or traits, among others. A common abstraction helps ensure a consistent data representation across all the supported languages and different components of the project, even though different languages may have unique constructs or terminology for similar entities. For example, classes in Python are represented using the similar data model abstraction as types in other languages. This consistency makes it easier for developers and tools to understand and manipulate code, regardless of the underlying language. 

% Introduce a notion of adaptors that take the information generated by backend tool and ensure it follows the language specific data model.

\cldk uses language-specific adapters that transforms artifacts generated by the backend analysis into a consistent data schema based on pydantic. 
The pydantic notations allows for diverse constructs (e.g., modules in Python, packages in Java) to be represented uniformly according to the schema. This also ensures that the APIs offered to developers are consistent across languages, enabling the same set of methods and queries to be used irrespective of the programming language.
\begin{figure}[t!]
\begin{code}
\begin{minipage}[t]{0.5\linewidth}
\vspace{-0.5em}
\begin{minted}[linenos=false, baselinestretch=1]{python}
|\textbf{class}{\color{blue} \textbf{PyField}}:|
  field_name: str 
  field_type: str
  line_offset: List[int, int]

|\textbf{class}{\color{blue} \textbf{PyArg}}:|
  arg_name: str
  arg_type: str 

|\textbf{class}{\color{blue} \textbf{PyClass}}:|
  code_body: str
  file_name: Path
  fields: |List[\hlblue{PyField}]\labeled{7.8}{a}|
  full_signature: str
  class_name: str = None
  super_classes: List[str]
  callables: |List[\hlblue{PyCallable}]|
  |\hspace{13.4em}\rtanglelabeled{7}{b}|
  line_offset: Tuple[int,int]

|\textbf{class}{\color{blue} \textbf{PyCallable}}:|
  code_body: str
  is_static: bool
  method_name: str
  full_signature: str
  is_constructor: bool
  formal_params: |List[\hlblue{PyArg}]\labeled{2.255}{c}|
  callsite: |\scriptsize{List[\hlblue{PyCallSite}]\labeled{2.255}{d}}|
  line_offset: Tuple[int,int]
  return_type: str |$\mathbf{\vert}$| PyClass
  
\end{minted}
\vspace{-2.5em}
\vspace{0.5em}
\label{listing:cldk-core}
\end{minipage}%
\begin{minipage}[t]{0.5\linewidth}
\vspace{-0.5em}
\begin{minted}[linenos=false, baselinestretch=1.005]{python}
|\textbf{class}{\color{blue} \textbf{PyImport}}:|
  from: str 
  imports: List[str]

|\textbf{class}{\color{blue} \textbf{PyCallSite}}:|
  target_method: str
  arguments: List[str]
  col_offset: Tuple[int, int]
  line_offset: Tuple[int, int]

|\textbf{class}{\color{blue} \textbf{PyModule}}:|
  file_name: Path
  qualified_name: str
  classes: |List[\hlblue{PyClass}]\labeled{6.9}{e}|
  imports: |List[\hlblue{PyImport}]\labeled{5.75}{f}|
  callables: |List[\hlblue{PyCallable}]|
  |\hspace{13.4em}\rtanglelabeled{7.}{b}|
|\textbf{class}{\color{blue} \textbf{PyBuildAttributes}}:|
  build_file_type: str
  build_tool: str
  package_name: str = None
  version: str = None
  dependencies: List[str]
  scripts: List[str]

|\textbf{class}{\color{blue} \textbf{PyConfig}}:|
  config_file_name: str
  code_body: str  
  config_type: str 
  settings: Dict[str, Any]
\end{minted}
\end{minipage}
\vspace{-1em}
\captionof{listing}{A sample schema for python applications.}
\label{listing:cldk-python-schema}
\end{code}
\end{figure}

\snippet{cldk-python-schema} shows a (simplified) version of the pydantic schema for analyzing python projects. This schema is organized into multiple classes that contains a native python class (that extends a pydantic \inlinecode{BaseModel}) to encapsulate various components of Python applications such as imports (via \inlinecode{PyImport}), fields (via \inlinecode{PyField}), modules (via \inlinecode{PyModule}), classes (via \inlinecode{PyClass}), etc. The fields of the classes are either native python types such as lists, tuples, or strings, or they reference other classes in the schema. For instance, \inlinecode{PyModule} and \inlinecode{PyClass} contain a field (\inlinecode{callables}) which is an array of \inlinecode{PyCallable} object (see \circled{b} in~\snippet{cldk-python-schema}). Further, \inlinecode{PyCallable} contains fields \inlinecode{formal_params} and \inlinecode{callsite} that are of type \inlinecode{List[PyArg]} (\circled{c},~\snippet{cldk-python-schema}) and \inlinecode{PyCallSite} (\circled{d},~\snippet{cldk-python-schema}), respectively.

This interconnection eases the development of APIs over the schema that perform detailed analysis of Python code. The analysis depth could vary from simple tasks like listing all the imports in a module, to such complex tasks as understanding classes and their members (fields and callables), inspecting function signatures and arguments, identifying call sites, tracking imports, and managing modules and configurations. To highlight this, in~\snippet{python-get-callables}, we demonstrate a simple API implementation for python applications to get all the callable definitions in a python module. Such an enumeration is often useful for such tasks as code completion, test generation, generating module summaries, among others. We observe that the application schema for python (defined in~\snippet{cldk-python-schema}) makes it easy to define API in just a few lines of code. Further, fields such as \inlinecode{is_constructor}, defined over \inlinecode{PyCallable} allows us to extend the logic with filter commands to offer more precise outputs (see \circled{a},~\snippet{python-get-callables}).

\begin{figure}[t!]
\begin{code}
\vspace{-0.5em}
\begin{minted}{python}
# Get all the imports in a module
class PyAnalysis:
  @classmethod
  def get_all_methods(cls, module: PyModule) -> List[PyCallable]:
    """List all the callable methods in a module
    Args:
      module (PyModule): The module in which to find imports in.
    Returns:
      List[PyCallable]: List of all the methods."""
    methods = module.callables  # Gather callables in module
    methods.extend(|\hlblue{filter}(\labeled{43.4}{a}|
      lambda meth: not |\hlblue{meth.is\_constructor}, \labeled{22.5}{a}|
        klass.callables))
    return methods
\end{minted}
\vspace{-2.5em}
\captionof{listing}{A sample API call over a python module to get all the methods inside it (that are not constructors).}
\label{listing:python-get-callables}
\end{code}
\end{figure}

\subsubsection{LLM Interaction}

Program analysis artifacts must often be supplied to Code LLMs in the form of non-standard, highly structured formats such as graphs or trees; these appear adjacent to natural language text and code blocks. Formulating such prompts (not only for interacting with models, but also to instruction tune models) is a challenging problem often highlighted by developers. There exist several prompting, composing, and templating tools such as LangChain~\cite{langchain}, LMQL~\cite{lmql} and/or Guidance~\cite{guidance}. However, these tools seldom offer code-specific plugins, and it is non-trivial to extend them to easily integrate with program analysis context. Templating tools native to python such as Jinja~\cite{jinja} do not easily generalize to new use-cases without considerable amount of developer effort. Further, they are challenging to build, test, and verify at enterprise scale, especially when tackling complex analysis artifacts or code-constructs.

\cldk tackles these issues by providing a code-specific prompt composition API that is tightly coupled with the Pydantic schema discussed above. This integration ensures that all data used in prompt composition is well-typed, validated, and directly reflects the code's structure and semantics without much overhead on the developers. This integration is best demonstrated in~\snippet{cldk-prompting-api} which contains a (contrived) example of implementing a unit test generation for all methods in a python project using \cldk. 

\begin{figure}[tb!]
\begin{code}
\vspace{-0.5em}
\begin{minted}{python}
from cldk import CLDK, Languages
from cldk.prompting import Composer, PromptBuilder
from cldk.schema import python
# Intialize CLDK
cldk = CLDK(language=Languages.Python)}
analysis = cldk.analysis(project\_path=Path(...)}

for module in analysis.get_all_modules():
  for callable in analysis.get_all_methods(module=module):
    method_call_graph = analysis.get_method_call_graph(callable)
    skeleton = |\hlblue{Composer()}\labeled{44.5}{a}|
    skeleton.add_line(
    "Generate a pytest unit test for the method")
    skeleton.add_line(
      "{callable.method_name} in {module.qualified_name}.")
    skeleton.|\hlblue{add\_code\_block}|(|\labeled{41}{b}|
      "{callable.code_body}", lanugage="python")
    skeleton.add_line(
      "Given the following contextual information:")
    skeleton.add_graph(|\hlblue{method\_call\_graph}|)|\labeled{25.8}{c}|
    # Build the prompt using CLDK's prompting module
    prompt = PromptBuilder(|\hlblue{provider}|="langchain")|\labeled{5}{d}|
      .template(schema=python.schema, skeleton=skeleton)
    # Execute the prompt using ollama
    result = prompt.ollama.run(model_id="granite-8b-code")
\end{minted}
\vspace{-2.5em}
\captionof{listing}{\cldk APIs make it trivially straightforward to accomplish SE tasks (unit test generation in this case).}
\label{listing:cldk-prompting-api}
\end{code}
\end{figure}
\section{Case Studies of \cldk in watson{\color{blue}x} Code Assistant}
\label{sect:industrial-usecases}
In this section, we expand the usefulness and the productivity impact of \cldk on two use cases -- (a) test generation and (b) code explanation. Both features are part of IBM watsonx Code Assistant product. First, we discuss the high-level overview of both use cases and then show various examples of using \cldk to enable static analysis, prompting, and performing post-processing on LLM output.

\subsection{Unit-test generation}
\label{sect:aster}
\begin{figure}[htp]
    \centering
    \includegraphics[width=\linewidth]{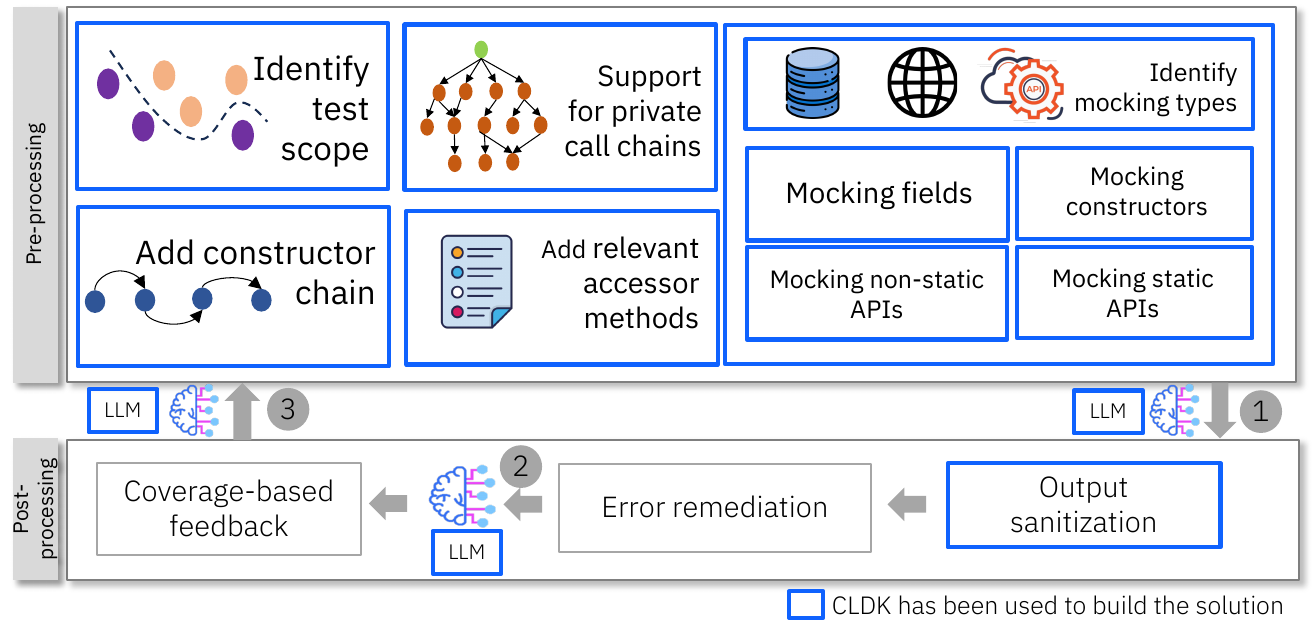}
    \caption{Architectural diagram of test generation capability.}
    \label{fig:asteroverview}
    %\vspace{-10pt}
\end{figure}

Figure~\ref{fig:asteroverview} depicts the overall architecture of the automated approach of unit test generation using LLM. Broadly, the approach is divided into two phases - (a) pre-processing and (b) post-processing. During the pre-processing phase, using static analysis context related to the focal method has been obtained. In this section, we will discuss few use cases and how CLDK has helped us achieve that with ease.

\subsubsection{Pre-processing} During the pre-processing stage, we identify various contextual information that is required for generating test cases. Here, we will expand on a few of these pre-processing steps and explain how easily they can be incorporated using \cldk.

\begin{itemize}
    \item Identifying methods under scope. Let's assume given a qualified class name, we want to identify all the methods that can be treated as focal methods, which are all the methods other than private methods. While using reflection, the private method can be accessed, but that is not a recommended practice. Below, we show how 4 lines of code can achieve that. First, we call \texttt{get\_methods} to gather all the methods given the qualified class name. Then, we go through each method and select the ones that are not private.
    \begin{minted}[frame=lines,framesep=1mm,baselinestretch=0.5, fontsize=\scriptsize, breaklines, breakanywhere, linenos,numbersep=2pt, mathescape=true, escapeinside=||]{py}
 def __init__():
    cldk = CLDK(language="java")
    self.analysis = cldk.analysis(project_path=<path>, analysis_level=AnalysisLevel.call_graph)
    
  def identify_non_private_methods(self, class_name: str):
    method_to_test = []
    all_methods_in_class = self.analysis.get_methods(class_name)
    for method in all_methods_in_class:
        if 'private' not all_methods_in_class[method].modifiers:
            method_to_test.add(method)
    return method_to_test
\end{minted}
    \item \textbf{Identifying inherited classes.} To collect more context related to the classes extended in the focal class, we list all the classes in the inherited call chain. First, we get the focal class details and go through each extended member. Finally, recursively add class names.
     \begin{minted}[frame=lines,framesep=1mm,baselinestretch=0.5,  fontsize=\scriptsize, breaklines, breakanywhere, linenos,numbersep=2pt]{py}
    
  def get_inheritance_chain(class_name: str) -> List[str]:
    inheritance_chain = []
    class_details = analysis.get_class(class_name)
    inheritance_chain.extend(class_details.extends_list)
    for parent_class in class_details.extends_list:
        inheritance_chain.extend(get_inheritance_chain(parent_class))
    return inheritance_chain
\end{minted}
    \item \textbf{Generate private call graph.} Private methods cannot be directly tested, we therefore added support for private methods by adding the relevant details when a focal method calls a private method. For instance, let's assume we want to test \texttt{foo}, which calls a private method \texttt{bar}, which also calls another private method \texttt{baz}. To create a test case that covers \texttt{bar} and \texttt{baz}, LLM requires sufficient information surrounding the private call chain.
    % \begin{minted}[frame=lines,framesep=1mm,baselinestretch=0.5, fontsize=\scriptsize, breaklines, breakanywhere, linenos,numbersep=2pt]{java}
    % void foo(){
    %     bar();}
    % private void bar(){
    %     if (condtion)
    %         baz();}
    % private void baz(){...}
    % \end{minted}
    To implement the feature, one can use CLDK with less than 10 lines of code to get all the private methods in the call chain started from the focal method.  
    \begin{minted}[frame=lines,framesep=1mm,baselinestretch=0.5, fontsize=\scriptsize, breaklines, breakanywhere, linenos,numbersep=2pt]{py}
    
  def get_private_call_chain(class_name: str, method_signature: str, call_chain=[]):
    class_cg = analysis.get_class_call_graph(class_name, method_signature)
    for call in class_cg:
        target_method: JMethodDetail = call[1]
        if 'private' in target_method.method.modifiers:
            if target_method.method.signature not in call_chain:
            call_chain.add(target_method.method.signature)
            call_chain.extend(get_private_call_chain(class_name, target_method.method.signature, accessible_private_methods)
    return call_chain
\end{minted}
    % \item Generate constructor call chain
    \item \textbf{Identify mocking types.} To identify whether a focal method requires mocking, one can look at every call site and the receiver type of the call site and decide whether the receiver type falls under the predefined set of classes that the user wants to enable mocking for. This implementation can be extended for more complicated cases as we discussed in \cite{pan2024multi}. 
        \begin{minted}[frame=lines,framesep=1mm,baselinestretch=0.5, fontsize=\scriptsize, breaklines, breakanywhere, linenos,numbersep=2pt]{py}
    
  def is_mocking_type(class_name: str, method_signature: str) -> List[str]:
    method_details = analysis.get_method(class_name, method_signature)
    for call_site in method_details.call_sites:
        if call_site.receiver_type in mockable_types:
            return True
    return False
\end{minted}
\end{itemize}
\subsubsection{Calling LLMs}
In this step, we use \texttt{Jinja} template alongwith \cldk API to communicate with LLMs hosted in (a) local server and (b) cloud service (either through http protocol or API).
\subsubsection{Post-processing}
LLMs are prone to making mistakes and require guardrails to rectify. Specifically in case generating unit test for Java, this step involves (a) parsing the output, (b) fixing simple mistakes, e.g., not adding required \texttt{@Test} annotation, unmatched parenthesis, (c) adding imports and package details, etc. On top of that, compilation errors and runtime errors can be identified, and the error-related details can be used for feedback to LLM, which can attempt to fix the mistakes. In this entire pipeline, \cldk has been used in (a) code parsing and (b) generating prompts. Using \cldk, one can use \texttt{is\_parsing(language:str)} method that returns the parsing errors by running Tree-sitter on the code (or snippet).

\subsection{Code Summarization}
\label{sect:summarization}
Program analysis has proven to be an essential technique for improving the accuracy and effectiveness of code summarization—the task of generating natural language descriptions of source code. By leveraging program analysis, code summarization models can move beyond producing surface-level descriptions based solely on keywords or comments and provide summaries that capture deeper aspects of code structure, logic, and behavior. \cldk ability to identify variable usage, function dependencies, and control structures enables to identify vital code elements and their contributions to functional logic that can be easily incorporated into prompts irrespective of the model and programming language. One such use case can be introducing the \texttt{callers} and \texttt{callees} in to code context to generate more meaningful explanation of the code. Using \cldk API, one can easily gather such information.

% \begin{figure}[t!]
% \begin{code}
% \vspace{-0.5em}
% \begin{minted}{java}
% %  public OrderDataBean sell(String userID, int holdingID, int orderProcessingMode) throws Exception {
% %         return sell(userID, new Integer(holdingID), orderProcessingMode);
% %     }

% % # Explanation:
% % TradeAction.sell method has three input parameters:
% % 1. userID: The ID of the user requesting the sale.
% % 2. holdingID: The identifier of the stock holding to be sold, provided as an Integer.
% % 3. orderProcessingMode: An integer representing the mode in which the order should be processed. 
% % It is reponsible for processing the sale of a stock holding for a specified user. It can be invoked from PortFolioJSF which coordinates the sale process by managing user sessions and holding data.
% % \end{minted}
% % \vspace{-2.5em}
% % \captionof{listing}{TradeAction.sell method from Daytrader and dependency aware brief explanation}
% % \vspace{-1em}
% % \label{listing:java-ts.sell}
% \end{code}
% \end{figure}

\label{sect:summarization}

% \begin{figure}[t!]
% \begin{code}
% \vspace{-0.5em}
% \begin{minted}{java}
%  callers = cldk.get_callers(class_name, prompt_dict["method_signature"],use_source_analysis)

%  callees = cldk.get_callees(class_name, prompt_dict["method_signature"], use_source_analysis)

%  def format_inst(code, focal_method, focal_class, language):
%  ...

%  def prompt_ollama(message: str, model_id: str = "granite-code:8b-instruct") -> str:
%  ..
 
% \end{minted}
% \vspace{-2.5em}
% \captionof{listing}{Steps to extract caller-callee information to create  enriched instruction for method explanation}
% \vspace{-1em}
% \label{listing:calleecaller}
% \end{code}
% \end{figure}

% \subsection{Productivity gain}
% \label{sect:productivity-gain}
% % Table generated by Excel2LaTeX from sheet 'Sheet1'
% \begin{table}[htbp]
%   \centering
%   \caption{Comparison of LOC needed to implement each use case with and without CLDK}
%   \resizebox{\columnwidth}{!}{  \begin{tabular}{l|r|r}
%     \multicolumn{1}{c|}{Task}  & \multicolumn{1}{l|}{LOC using CLDK} & \multicolumn{1}{l}{LOC w/o CLDK} \\
%     \hline
%     \hline
%     Identify testing scope &       &  \\
%     \hline
%     Support for private call chains &       &  \\
%     \hline
%     Add constructor call chain &       &  \\
%     \hline
%     Adding relevant accessor methods &       &  \\
%     \hline
%     Adding mocking-specific context &       &  \\
%     \hline
%     LLM calls &       &  \\
%     \hline
%     Output sanitization &       &  \\
%     \hline
%     \end{tabular}%
%     }
%   \label{tb:industryusecase}%
% \end{table}%

% \subsection{Lessons learned}

\section{Case Study on Academic Usecases}
\label{sec:acedemicusecase}

In this section we wanted to understand whether \cldk can help with the program analysis developed by some recent works on code LLM. To do that, we refer to our previously collected dataset and identify some potential examples where we \cldk can replace the existing implementation.

First, we manually went through each of these works and identified the particular usage of program analysis. Surprisingly, most of the works are on performing very light-weight program analysis such as getting AST, tokenizing AST, etc., However, a few of the works incorporated control flow graphs and data-flow graphs and extracted various parts of the code using program analysis. In this section, we will go through a few of these examples and describe how one could use \cldk to implement very similar functionality. One of the use case that we identified that validating whether the LLM-generated code can be parsed using the target language grammar, and mostly Tree-sitter has been used for that purpose. \cldk comes with \texttt{is\_parsing(language: str)} functionality that can be used off-the-shelf. Another interesting example, where the work~\cite{shrivastava2023repository} implemented extensive list of Tree-sitter queries to extract Java class name, class body, parent class name, method body, imports, field declarations, and other key parts of Java class. Below, we show a code snippet from the same work. First, in \texttt{get\_query} function, all the Tree-sitter queries are listed, which are processed in \texttt{get\_attribute} function. Then, there are specific functions for parsing the output of the Tree-sitter and getting information like import details, class name, parent file name, etc; in total, the analysis required 70+ lines of code.
\begin{code}
\begin{minted}[frame=lines,framesep=1mm,baselinestretch=0.5, fontsize=\scriptsize, breaklines, breakanywhere, linenos,numbersep=2pt, mathescape=true, escapeinside=||]{py}
def get_query(attribute_type):
  # list all the tree-sitter queries
  if attribute_type == 'class_name':
    query = JAVA_LANGUAGE.query("""(class_declaration
                            name: (identifier)@class_name)""")

  if attribute_type == 'class_body':
    query = JAVA_LANGUAGE.query("""(class_declaration
                            body:(class_body) @class_body)""")
  ...
  + 20 lines
def get_attribute(root_node, filename, attribute_type):
  # Processing tree-sitter queries
  + 6 lines of code
  return attributes

def get_parent_class_filename(parent_class_name, file_class_info, file):
  # get parent class file name
  + 18 lines of code
  return parent_class_filename


def get_imports(import_statements, file, all_identifiers, all_type_identifiers):
  # get imports
  + 28 lines of code
  return imports
\end{minted}
\end{code}

On the contrary, we show how easily a similar functionality can be implemented in 10 lines of code using \cldk. First, one needs to initialize \cldk object by providing the target programming language, which in this case is Java. Then, a user needs to provide the path to the application and use various off-the-shelf APIs like \texttt{get\_class}, \texttt{get\_methods\_in\_class}, etc., and other fields associated with each type for more details. 

\begin{code}
 \begin{minted}[frame=lines,framesep=1mm,baselinestretch=0.5, fontsize=\scriptsize, breaklines, breakanywhere, linenos,numbersep=2pt, mathescape=true, escapeinside=||]{py}
 def replace_with_cldk(class_name):
 
    # Initialize CLDK object
    cldk = CLDK(language="java")
    analysis = cldk.analysis(os.getenv("JAVA_APP_PATH"))
    
    # Get the class details
    class_details = analysis.get_class(class_name)
    class_name = class_name.split('.')[-1]
    
    # Read the class body
    with open(analysis.get_java_file(class_name), 'r') as f:
        class_body = f.read()
        
    # Get all the method details
    method_names = [method_name for method_name in analysis.get_methods_in_class(class_name)]
    method_bodies = [analysis.get_method(class_name, method_name).code for method_name in
                    analysis.get_methods_in_class(class_name)]
    
    # Get import and parent file details
    java_compilation_unit = analysis.get_java_compilation_unit(file_path=file_path)
    import_statements = ['import ' + imports + ';' for imports in java_compilation_unit.imports]
    parent_class_filenames = class_details.implements_list
\end{minted}
\end{code}

\section{Related Works}
\label{sec:related}
\cldk builds upon several established tools and frameworks, offering a unified interface that simplifies code analysis, prompting, and other essential features for LLM-based use cases. In this section, we highlight these foundational works and discuss the key distinctions.

\paragraph{Code Analysis} Code analysis has been a focus of research for decades, resulting in powerful tools like CodeQL~\cite{codeql}, Tree-sitter~\cite{treesitter}, Antlr~\cite{antlr}, WALA~\cite{wala}, Soot~\cite{soot}, Doop~\cite{doop}, Comex~\cite{das2023comex}, and others. Each of these frameworks differs in terms of the depth of analysis it performs, flexibility for custom analyses, and support for various programming languages. Furthermore, the expertise required to effectively integrate these tools can differ significantly. \cldk simplifies access to these static analysis tools by providing a unified wrapper, making it easier for developers to perform program analysis without needing in-depth knowledge of each tool's specifics. The key contributions of \cldk in this area include: (a) providing a unified framework that supports program analysis across multiple programming languages, (b) enabling different levels of program analysis, (c) abstracting away the complexity of underlying implementation details, and (d) creating a unified schema to extend the analysis to additional programming languages.

\paragraph{Prompting and Retrieval} Prompt engineering and orchestration play a critical role in interacting with LLMs. Prompt engineering and execution frameworks provide the ability to structure, control, and optimize complex workflows. \cldk leverages Jinja~\cite{jinja}, LangChain~\cite{langchain}, and Guidance~\cite{guidance} as templating engines and for managing prompt sequences and interactions with tools, respectively. LMQL~\cite{lmql}, etc, assist users in performing constraint-based decoding for more structured outputs and optimizing prompt blocks. To build retrieval pipelines that add relevant in-context examples, tools like ChromaDB~\cite{chromadb}, LlamaIndex~\cite{llamaindex}, and ElasticSearch~\cite{elasticsearch} facilitate integration into LLM-based applications. However, effectively integrating these disparate tools often requires considerable effort and expertise. In comparison, \cldk offers a unified wrapper around these tools, supporting a broad range of prompting and retrieval approaches and simplifying their integration. This unified approach reduces the learning curve and integration overhead associated with managing multiple prompting and retrieval tools, enabling developers to focus on enhancing the quality of LLM interactions.
\section{Limitations}
\label{sect:threats-to-validity}
This was conducted at IBM Research and while our research landscape provides a valuable foundation to ground our findings, we note that that some insights may be specific to the particular use cases and challenges being currently addressed. Thus, the findings may not be entirely generalizable to other organizations with different LLM applications.

The scope of IBM’s projects, however, and the wide range of research activities represent a significant portion of the population working with code LLMs, making our results indicative of the key challenges developers and researchers face when utilizing LLMs for program analysis and prompting.

Finally, \cldk is an evolving project. The tool’s capabilities, including language support and levels of analysis, are under active development. As the tool matures and expands to support more programming languages and deeper analysis features, the findings presented here should be viewed in the context of ongoing advancements in its development.
% \section{Discussion}
% \label{sect:discussion}

\section{Conclusion}
\label{sect:conclusion}
Code LLMs are used for tasks like code completion, generation, and test creation, but their effectiveness depends on providing rich, code-specific context—often derived from static analysis tools. These tools present challenges due to their steep learning curve and integration complexities, limiting their widespread use. To address this, we built \cldk as an open-source library that simplifies program analysis across multiple languages and frameworks to better contextualize Code LLM interaction. \cldk achieves this by providing an intuitive interface for developers to harness the full potential of LLMs more easily.
\IEEEtriggeratref{13}
\bibliographystyle{IEEEtran}
\bibliography{references}
\end{document}